# Beta oscillation changes in ALS: A Dual-Site International Replication Study

**Marit Boxum[#1], Gabriel Rodrigues Palma[#2], Robin Jansen[1], Iris A. Bastmeijer[1], Sanya Dalal[2], Eva Woods[2], Narin Suleyman[2,3], Eileen Rose Giglia[4], Aoife Troy[4], Abby Field[2], Serena Plaitano[4], Marjorie Metzger[4], Niall Pender[4,5], Orla Hardiman[3,4], Leonard H. van den Berg[1], Roisin McMackin*[2,4], Stefan Dukic*[1]**

[1]Department of Neurology, University Medical Center Utrecht Brain Center, Utrecht University, Utrecht, the Netherlands | [2]Discipline of Physiology, School of Medicine, Trinity College Dublin, University of Dublin, Dublin, Ireland| [3]Department of Neurology, Beaumont Hospital, Dublin, Ireland| [4]Academic Unit of Neurology, School of Medicine, Trinity College Dublin, University of Dublin, Dublin, Ireland| [5]Department of Psychology, Beaumont Hospital, Dublin, Ireland

[#]Joint first authorship
[*]Joint last authorship

E-mail: mcmackr@tcd.ie and S.Dukic@umcutrecht.nl

## Abstract

***Objectives.*** Objective neurophysiological biomarkers are needed to monitor motor and cognitive dysfunction in amyotrophic lateral sclerosis (ALS). Previous work has demonstrated reduced beta-band event-related desynchronization (ERD) and attenuated post-movement beta event-related synchronization (ERS) over frontal and parietal regions during the sustained attention to response task (SART) in individuals with ALS. We aimed to validate beta-band ERD and ERS alterations as potential biomarkers by replicating previous findings in independent Dutch and Irish cohorts. ***Approach***. A randomized SART with 128-channel electroencephalography (EEG) was performed in Dutch (ALS, $n$=63; age- and sex-matched controls, $n$=64) and Irish (ALS, $n$=36; age- and sex-matched controls, $n$=36) cohorts. Non-phase-locked oscillatory activity was quantified using a complex Morlet wavelet transform to compute baseline-normalized post-stimulus power decreases and increases corresponding to ERD and ERS. Group differences in ERD and ERS, and their associations with their task performance, were examined during Go (motor response) and NoGo (withhold response) trials. ***Main results***. Consistent with prior findings, the SART elicited theta-band ERS and alpha- and beta-band ERD, followed by rebound beta-band ERS across the frontoparietal axis during Go and NoGo trials in both control cohorts. Compared with Go trials, NoGo trials showed greater theta ERS, stronger alpha ERD, and attenuated beta-band ERS. In contrast to previous null findings, people with ALS showed reduced Go and total trial accuracy compared with controls in both centres ($p$<0.002) and slower Go response times in the Irish cohort ($p$=0.02). Compared with the Dutch ALS cohort, Irish people with ALS showed higher NoGo trial accuracy, slower Go responses, and fewer anticipation errors (all $p$=0.03). Consistent with prior findings, the Irish ALS cohort exhibited reduced beta-band ERD and ERS compared with controls, whereas no such significant group differences were observed in the Dutch cohort. Greater beta-band ERS was associated with faster Go response times in both centres ($p$<0.011), consistent with previous findings. ***Significance.*** ALS-associated ERD and ERS changes were replicated in the Irish, but not the Dutch cohort. Metrics which differed between ALS cohorts were associated with varying degrees of functional impairment. Reproducible control findings and the consistent associations with motor task performance further evidence reliable capture of cortical motor network dysfunction by these measures and demonstrate utility as an EEG-based marker of motor decline in ALS.

## 1. Introduction

Amyotrophic lateral sclerosis (ALS) is a progressive disorder characterized by degeneration of upper and lower motor neurons, leading to muscle weakness, respiratory failure, and ultimately death. Beyond these motor symptoms, up to 50% of people with ALS experience cognitive and/or behavioral impairments, with 15% meeting criteria for frontotemporal dementia [10, 14, 15].

The heterogeneous clinical presentation, variable progression, and overlap with other neurological conditions complicate both diagnosis and prognostication of ALS [5, 13]. Diagnosis relies on the identification of characteristic clinical features supported by neurophysiological and imaging investigations, and diagnostic delays remain common [9, 11]. While electromyography provides established biomarkers of lower motor neuron dysfunction [39], robust biomarkers of upper motor neuron impairment remain elusive. This limits the ability to quantify disease-related neurophysiological changes and to sensitively monitor disease progression or treatment-related effects. Neurophysiological biomarkers that reflect cortical network dysfunction could complement existing clinical assessments by providing objective measures of disease activity and treatment response, as well as outcome measures for clinical trials [3, 12].

Electroencephalography (EEG) offers several advantages for biomarker discovery compared with other neuroimaging modalities [3]. Its relatively low equipment and operational costs and widespread availability make it an accessible tool for both routine screening and longitudinal monitoring. Moreover, EEG provides a direct measure of neural network function with high temporal resolution, unlike indirect measures such as blood oxygenation levels or metabolic activity, which can be influenced by age- and disease-related factors [6]. Furthermore, recordings can be obtained while participants are seated, enhancing comfort and feasibility in ALS [4]. Taken together, these features make EEG a promising tool for investigating brain networks in ALS.

A number of groups have investigated the potential of EEG to identify patterns of brain network dysfunction that can detect and measure ALS pathophysiology. These patterns involve oscillations of specific frequencies, typically grouped into frequency bands: delta (<4 Hz), theta (4–7 Hz), alpha (8–12 Hz), beta (13–30 Hz), and gamma (>30 Hz). Beta-band event-related desynchronization (ERD) and synchronization (ERS), commonly associated with motor preparation and execution, and post-movement processing, respectively, have been proposed as promising functional biomarkers in ALS. Reduced beta-band ERS has been proposed to reflect impaired inhibitory control over upper motor neurons in ALS [3]. EEG studies using a range of motor paradigms, including self-paced movements, have demonstrated impaired beta-band ERD and ERS in ALS, supporting their potential utility as markers of cortical motor dysfunction [40,41]. McMackin et al. [8] previously demonstrated that beta-band ERD and ERS during the sustained attention to response task (SART), a Go/NoGo paradigm assessing sustained attention and inhibitory control, are diminished in ALS and correlate with task performance. Importantly, the SART incorporates both motor and cognitive demands, enabling simultaneous assessment of motor and broader cognitive network dysfunction. In their cohort, ERD demonstrated good discriminative performance, with reported area under the receiver operating characteristic curve (AUROC) values of 0.80 (prefrontal; Fz) and 0.82 (parietal; Pz). Furthermore, controls showing higher beta-band ERS during Go trials performed the task quicker but less accurately, suggesting a speed–accuracy trade-off. When task performance metrics were compared, people with ALS exhibited only increased anticipation error rates, but not significant speed or accuracy differences, compared to controls.

In this paper, we aim to validate our prior findings regarding beta-band oscillations during the SART across two independent, multi-center cohorts recruited in Ireland and the Netherlands. Specifically, we compare the behavioral measures with event-related spectral perturbation (ERSP) responses from Go, NoGo, and NoGo-Go (the difference between NoGo and Go) trials in people with ALS and controls. Furthermore, we assess the diagnostic utility of beta-band ERD and ERS to distinguish people with ALS from controls. By evaluating these neurophysiological signatures across distinct international populations, we aim to establish the reproducibility and clinical scalability of beta-band SART ERSP as robust biomarkers in ALS.

## 2. Methods

### 2.1 Participants

People with ALS were recruited from the Motor Neuron Disease Outpatient Clinic of the University Medical Center (UMC) Utrecht and the National ALS clinic at Beaumont Hospital Dublin. Controls comprised neurologically healthy, age- and sex-matched individuals recruited from population-based datasets in the Netherlands and Ireland. All participants were aged 18 years or older, and provided written informed consent in accordance with the Declaration of Helsinki, with the exception of pre-registration of the study.

### 2.2 Experimental paradigm

The SART was administered following the experimental paradigm described in previous studies [4, 15, 19]. In short, participants were instructed to press the left mouse button each time a digit (from one to nine) appeared on the black screen (“Go trials”) while withholding responses to the number “3” ("NoGo trials"). Stimuli were displayed in randomized order for 250 ms using Presentation software (Neurobehavioral Systems Inc., Albany, CA), followed by an inter-stimulus interval of 1120-1220 ms. Digits were shown in light grey with randomized font sizes (100-180 points) to prevent perceptual templating. Participants were seated approximately 1 m from a computer monitor in a darkened room to minimize visual distractions. The task consisted of 5-minute blocks, each containing 252 trials, with the number "3" appearing randomly in 11% of trials. Participants were instructed to prioritize both speed and accuracy and completed three or four blocks of the task. This variation in block count resulted from protocol optimization aimed at reducing overall study duration, as an interim analysis indicated that the additional block yielded no substantial improvement in the signal-to-noise ratio. Before recording, participants completed a supervised practice round of up to 45 trials to ensure task comprehension.

**2.3 Data acquisition**

Task performance of the SART was recorded using Presentation software, while EEG data were acquired using a high-density 128-electrode BioSemi Active Two system (BioSemi B.V., Amsterdam, The Netherlands) at a sampling rate of 512 Hz with hardware-based lowpass filtering at 104 Hz to prevent aliasing.

Disease severity was assessed using revised ALS Functional Rating Scale (ALSFRS-R) scores [2]. Cognitive status was evaluated using the Edinburgh Cognitive and Behavioral ALS Screen (ECAS) [1], a screening tool designed to detect cognitive and behavioral changes in ALS, which was administered in validated Dutch or English versions to the respective cohorts. Data collected within 90 days before or after the EEG recording were included in the analyses. ECAS data were available for 20 (55%) people with ALS in the Irish cohort and for 58 (92%) people with ALS and 61 (97%) healthy controls in the Dutch cohort. ALSFRS-R scores were available for 30 (83%) and 62 (98%) people with ALS in the Irish and Dutch cohorts, respectively. King's clinical stage was estimated from the available clinical information according to the King's clinical staging system for ALS, which is based on the extent of regional involvement and the presence of significant nutritional or respiratory failure [38].

**2.4 Data Analysis**

*Task performance*

Behavioral data collection included: median response time, NoGo accuracy (percentage of three-digit stimuli followed by response omission), Go accuracy (percentage of nonthree-digit stimuli followed by a response within 170-650 ms), total accuracy (using both Go and NoGo trials), and anticipation error (response made within 170 ms of a Go stimulus). The adjustment of the anticipation threshold from 150 ms to 170 ms was implemented to better align with early stages of perceptual processing [36], indicating that responses occurring prior to this interval may not reflect fully processed perceptual information.

*EEG Data*

The manual data cleaning steps utilized in our previous work [8] were replaced with a fully automated EEG preprocessing pipeline, which has been described in detail elsewhere [35]. The automated pipeline applies a 0.3Hz dual-pass $4^{th}$ order Butterworth high-pass filter and resamples the data from 512Hz to 256 Hz. Flat and noisy electrode signals were then automatically detected and replaced using spline interpolation of surrounding channels, after which the data were re-referenced using a common average reference. Independent component analysis was subsequently performed to identify and remove stereotyped noise components, including ocular, muscle, and cardiac artefacts. The cleaned data were then low-pass filtered at 60 Hz using a dual-pass $4^{th}$ order Butterworth filter. Finally, the continuous EEG was segmented into epochs and baseline corrected.

In contrast to our previous preprocessing approach [8], which used shorter epochs optimized for time-domain event-related potential (ERP) analysis (e.g., N2/P3 analyses), the current preprocessing pipeline was adapted to accommodate time frequency analyses as well. Specifically, wider epochs (1100 ms pre-stimulus to 1100 ms post-stimulus) were employed to reduce edge effects within the time window of interest and to increase the temporal window available for baseline power estimation. Additionally, the use of a higher low-pass filter was intended to better preserve oscillatory activity relevant for time-frequency decomposition. After preprocessing, the data were visually screened, and participants with excessive noise or artifact-contaminated data were excluded from further analysis.

Processed EEG data were validated as comparable to those processed via the pipeline employed in our previous paper [8]. The validation was assessed through visual inspection of the averaged waveforms alongside descriptive statistics of N2/P3 peak amplitude and latency [7]. Additionally, Spearman's rank correlations were computed between the ERP components derived from each pipeline for comparison. Overall, the two pipelines showed strong agreement across both ERP components. For the N2 component, peak amplitudes were highly correlated between pipelines (mean Spearman's $\rho = 0.94$), whereas latency correlations were slightly lower (mean Spearman's $\rho = 0.79$). The P3 component showed a similar pattern, with high peak amplitude correlations (mean Spearman's $\rho = 0.93$) and slightly lower latency correlations (mean Spearman's $\rho = 0.83$). All correlations were statistically significant ($p < 0.01$). Detailed results are provided in Supplementary Material, Section 3.

*Time-frequency analysis*

Each trial epoch consisted of 2200 ms of data per channel at a sampling frequency of 256 Hz. As in the previous publication, a random subset of Go trials with correct responses was matched to the number of NoGo trials with correct responses to achieve equal numbers of trials per condition in the analyses. For each participant, matched trials were selected by drawing a random permutation of trial indices to obtain equal numbers of trials across conditions, thereby reducing potential imbalances in trial counts between conditions. Time-frequency decomposition was performed using complex Morlet wavelets with the bandwidth parameter (b0) set to 1, providing the optimal balance between temporal and spectral resolution. In contrast to our previous study, data padding at the beginning and end of epochs was not required due to the increased epoch length. Since the data were low-pass filtered at 60 Hz (compared to 35 Hz in our previous study), wavelet amplitudes were calculated over a broader frequency range (1–50 Hz), which was not feasible in the previous study due to the lower cutoff frequency. Mean inter-trial phase variance (ITV) was computed across epochs (e) for each time point (t) and frequency (f) as a measure of oscillatory activity that was not phase-locked to the stimulus:

$$ITV(f,t) = \frac{\sum_{1}^{N_e} \left| \left(W_{f,t,e} - \overline{W_{f,t}}\right)^2 \right|}{N_e - 1}$$

where $\overline{W}$ denotes mean value of W across all epochs and $N_e$ the total number of epochs. ITV was calculated for the Fz, Cz and Pz electrodes, previously selected for their proximity to cortical regions associated with executive and motor functions. Baseline values for each frequency were calculated as mean ITV within the −200 ms to 0 ms pre-stimulus interval. ERSP values were subsequently calculated for each time point and frequency as:

$$ERSP(f,t) = 100\ x\ \frac{ITV_{f,t} - \overline{ITV_{f,baseline}}}{\overline{ITV_{f,baseline}}}$$

for t = 1–1100 ms. As in our prior study [8], the temporal dimension of the time–frequency representations was downsampled to 34 Hz (i.e., every 15th time point was retained) prior to statistical analysis. This reduced computational load and the number of statistical comparisons while preserving the frequency range of interest.

*Regions of Interest*

To investigate the reproducibility of our prior findings, mean ITV was calculated within time-frequency regions of interest (ROIs) identical to those used in our prior analysis [8]. These mean values (without down sampling) were subsequently used for correlation analyses with behavioral measures.

*Statistical analysis*

To maximise comparability, statistical tests employed are mostly identical to those implemented in our prior investigation of SART-associated beta ERSP in ALS.

Differences in demographics between groups were assessed using the Mann–Whitney U-tests for continuous variables and chi-squared proportion testing for categorical variables, with significance set at p = 0.05.

Group differences in task performance were assessed using Mann–Whitney U tests, with. *p*-values adjusted for multiple comparisons using the Benjamini–Hochberg false discovery rate (FDR) correction (q = 0.05).

Significant ERSPs in controls (i.e., values differing from zero) were assessed using Wilcoxon signed-rank tests, while group differences in ERSPs were evaluated using AUROC. Following the statistical approach described in our previous work [8], empirical Bayesian inference was used to estimate posterior probabilities and statistical power, while multiple-comparison corrections were controlled via FDR (q = 0.05). To facilitate interpretation, AUROC values below 0.5 were transformed as *1-|AUROC - 0.5|*, such that larger values consistently reflected stronger discrimination between groups irrespective of whether oscillatory activity increased or reduced in people with ALS, given that the AUROC is calculated based on ERS and ERD. To complement the AUROC analyses and further characterize the potential clinical utility of ERSP measures, sensitivity and specificity measures are reported in the Supplementary Material, Section 1.

Correlations between ERSP measures and task performance were assessed using Spearman's rank correlation coefficients. Previously reported significant associations between beta-band ERS and task performance were examined for replication [8].

Due to deviations in findings between centres, additional analyses were performed to investigate associations between beta-band ERSP measures and clinical heterogeneity. Specifically, associations between beta-band ERD and task performance were assessed using Spearman's rank correlation coefficients for measures that differed between the Dutch and Irish ALS cohorts. For these analyses, electrodes of interest (Fz and Pz) were defined based on findings in the current study and a previous study [8]. *P*-values were corrected for multiple comparisons using FDR correction. Additionally, linear models were employed to determine if clinical factors that differed between ALS cohorts related to beta-band ERSP measures. Separate linear models were fitted, one for each combination of measure, electrode and trial type: ERS at Fz, ERS at Pz, ERD at Fz, and ERD at Pz. Each model included the effects of King's Stage, Riluzole usage, Country, and education level (high or low) as predictors. Goodness-of-fit was assessed using half-normal plots [37], and the significance of the effects was evaluated using F tests.

In addition, to assess whether between-cohort differences in beta-band ERSP measures could be driven by differences in the number of trials used for time–frequency analysis, the number of trials was compared between people with ALS and controls within each cohort and between the pooled Irish and pooled Dutch cohorts using Mann–Whitney U tests.

To assess whether the observed findings were dependent on the choice of event-locking procedure, ERSP analyses were also performed using response-locked epochs. Visual inspection indicated similar ERSP topographies and temporal patterns to those observed in the stimulus-locked analyses.

## 3. Results

### 3.1 Participant demographics

Table 1 presents the demographics and clinical characteristics of the cohorts from both centres. The Dutch cohort included 63 people with ALS and 63 controls, after exclusion of 12 people with ALS and 1 control due to insufficient data quality (n = 7), incorrect task execution (n = 1) or factors that could confound electrophysiological findings (n = 5), including a history of significant head trauma or active psychoactive medication use. The Irish cohort included 36 people with ALS and 36 controls. Both centres had cohorts that were age- and sex-matched.

Significant differences between cohorts were found in control age (median [interquartile range]: Irish 60 [41–67] vs. Dutch 65 [56–71] years; p = 0.01), control sex (male: Irish 19 [52.8%] vs. Dutch 45 [71.4%]), and educational level among people with ALS (high education: Irish 26 [72.2%] vs. Dutch 32 [50.8%]; p = 0.005). Among people with ALS, King's stage (p = 0.03) and riluzole use (p = 0.02) also differed between cohorts. Riluzole use was more prevalent in the Irish cohort [86.1%] compared to the Dutch cohort [69.8%], with median durations of 5.6 and 4.1 months, respectively. For King's stage, the Irish cohort showed a higher proportion of people with ALS in stages 1 [42.8% vs 23.8%] and 4 [25.0% vs 11.1%], whereas the Dutch cohort showed higher proportions in stages 2 [25.0 vs 36.5%] and 3 [7.1% vs 27.0%].

### 3.2 Task performance

Table 2 presents the summary of the SART performance metrics. In the Dutch cohort, Go accuracy was lower in people with ALS (96.3% [92.7–98.4]) compared to controls (97.8% [96.5–99.1]; p = 0.002). Moreover, total accuracy was lower in those with ALS (92.7% [89.5–96.0]) compared to controls (95.3% [93.7–97.0]; p = 0.002), while anticipation error was higher in people with ALS (0.9 [0.3–3.3]) compared to controls (0.7 [0.1–1.5]; p = 0.04).

**Table 1.** Demographics and clinical characteristics of the Irish and Dutch cohorts for people with amyotrophic lateral sclerosis (ALS) and controls. Data are shown as count [%] and medians [interquartile ranges].

| | **Dutch** | | | **Irish** | | | **Dutch vs Irish** | |
|---|---|---|---|---|---|---|---|---|
| | ALS (*n*=63[a]) | Controls (*n*=63) | *p* | ALS (*n*=36[b]) | Controls (*n*=36[c]) | *p* | ALS, *p* | Controls, *p* |
| *Age at first EEG, years* | 65 [58-71] | 65 [56-71] | 0.66 | 63 [58-68] | 60 [41-67] | 0.13 | 0.21 | 0.01 |
| *Sex, male* | 47 [74.6] | 45 [71.4] | 0.84 | 25 [69.4] | 19 [52.8] | 0.23 | 0.75 | 0.046 |
| *Handedness, right* | 56 [88.9] | 56 [88.9] | 0.76 | 32 [88.9] | 32 [88.9] | 1.00 | 0.61 | 0.81 |
| *Education level, high* | 32 [50.8] | 39 [61.9] | 0.21 | 26 [72.2] | - | - | 0.005 | - |
| *ECAS, abnormal total score* | 2 [3.1] | 0 [0] | 0.24 | 2 [5.6] | - | - | 0.26 | - |
| *ECAS, abnormal ALS-specific score* | 3 [4.7] | 0 [0] | 0.11 | 3 [8.3] | - | - | 0.18 | - |
| *ECAS, abnormal ALS-nonspecific score* | 0 [0] | 0 [0] | - | 1 [2.8] | - | - | 0.26 | - |
| *Disease duration, months* | 18.6 [11.6-38.3] | - | - | 19.1 [13.8–28.0] | - | - | 0.92 | - |
| *EEG delay, months* | 5.2 [3.6-13.6] | - | - | 6.9 [4.2-11.0] | - | - | 0.33 | - |
| *Site of onset, spinal* | 44 [69.8] | - | - | 27 [75.0] | - | - | 0.82 | - |
| *C9orf72 repeat length expansion* | 6 [9.4] | - | - | 1 [2.8] | - | - | 1.00 | - |
| *King's stage, 1/2/3/4* | 15/23/17/7 [23.8/36.5/27.0/11.1] | - | - | 12/7/2/7 [42.8/25.0/7.1/25.0] | - | - | 0.03 | - |
| *ALSFRS-R total score* | 40.0 [36.0-42.0] | - | - | 41 [38.0-43.0] | - | - | 0.09 | - |
| *Progression rate* | 0.5 [0.2-0.8] | - | - | 0.3 [0.2-0.6] | - | - | 0.36 | - |
| *Riluzole usage, yes* | 44 [69.8] | - | - | 31 [86.1] | - | - | 0.02 | - |
| *Riluzole duration, months* | 4.1 [3.3-6.6] | - | - | 5.6 [3.5-10.4] | - | - | 0.36 | - |

*P*-values were calculated using the Mann-Whitney U test for continuous variables and the chi-squared proportion testing for categorical variables.
[a] Missing data: Handedness (n=1), Education (n=1), ECAS total (n=2), ECAS specific (n=2), ECAS non-specific (n=2), King's stage (n=1), ALSFRS-R total (n=2), Progression rate (n=2).
[b] Missing data: Handedness (n=1), Education (n=5), ECAS total (n=12), ECAS specific (n=11), ECAS non-specific (n=11), Disease duration (n=1), EEG delay (n=1), C9orf72 (n=16), King's stage (n=8), ALSFRS-R total (n=6), Progression rate (n=6), Riluzole usage (n=2), Riluzole duration (n=5).

[c] Missing data: Handedness (n=2).
Abbreviations: ALS, amyotrophic lateral sclerosis; ECAS, Edinburgh Cognitive and Behavioral ALS Screen; EEG, electroencephalography; ALSFRS-R, revised ALS Functional Rating Scale.

**Table 2. Summary of task performance.** Data are presented as median [interquartile range].

| | Dutch | | | Irish | | | Dutch vs Irish | |
|---|---|---|---|---|---|---|---|---|
| | ALS (*n*=63) | Controls (*n*=63) | *p* | ALS (*n*=35[a]) | Controls (*n*=36) | *p* | ALS, *p* | Controls, *p* |
| *Go accuracy (%)* | 96.3 [92.7-98.4] | 97.8 [96.6-99.1] | 0.001 | 94.1 [89.7-98.1] | 98.4 [96.8-99.3] | $9.47*10^{-4}$ | 0.47 | 0.56 |
| *NoGo accuracy (%)* | 76.2 [59.3-81.5] | 79.3 [67.3-85.7] | 0.09 | 81.1 [72.1-88.7] | 83.0 [73.1-90.1] | 0.58 | 0.03 | 0.26 |
| *Total accuracy (%)* | 92.7 [89.5-96.0] | 95.4 [93.7-97.0] | 0.001 | 92.4 [88.2-95.7] | 96.0 [94.1-97.5] | $9.47*10^{-4}$ | 0.32 | 0.56 |
| *Go response time (ms)* | 327.8 [286.0-366.2] | 319.3 [294.1-342.7] | 0.58 | 360.5 [314.9-392.0] | 324.2 [290.8-359.0] | 0.02 | 0.03 | 0.67 |
| *Go anticipation error (%)* | 0.9 [0.3-3.3] | 0.7 [0.1-1.5] | 0.04 | 0.3 [0.1-1.2] | 0.2 [0.1-0.7] | 0.44 | 0.03 | 0.26 |

*P*-values were calculated using the Mann–Whitney U test with Benjamini–Hochberg FDR correction (q=0.05).
[a] SART performance data was missing for one patient.
Abbreviations: ALS, amyotrophic lateral sclerosis; FDR, false discovery rate; SART, sustained attention to response task.

Similarly, in the Irish cohort, Go accuracy was lower in people with ALS (94.1% [89.7–98.1]) compared to controls (98.4% [96.8–99.3]; $p < 0.001$). Total accuracy was also lower in the ALS group (92.4% [88.2–95.7]) than in controls (96.0% [94.1–97.5]; $p < 0.001$). Go trial response times were longer in people with ALS (median 360.5 ms [314.9–392.0]) compared to controls (324.2 ms [290.8–359.0]; $p = 0.02$).

Between-cohort comparisons revealed that people with ALS in the Irish cohort exhibited longer Go response times, higher NoGo accuracy, and lower anticipation error rates compared to those in the Dutch cohort (all $p = 0.03$), with no differences observed in Go accuracy or total accuracy.

### 3.3 Event-related spectral perturbations

Figure 1 illustrates the time–frequency ERSP patterns observed in the Dutch and Irish control cohorts during Go and NoGo trials, and their overlap with our previously defined time-frequency ROIs [8].

*Go Trial*

The ERSP observed in the Dutch and Irish cohorts during Go trials is presented in Figure 1A. In both centres, ERS (i.e., increase in power relative to baseline) was observed in the theta-band (4–7 Hz) and ERD (i.e., decrease in power relative to baseline) in the alpha- (8–12 Hz), and beta- (13–30 Hz) bands across the frontoparietal axis. In the beta-band, ERS followed ERD and was significant across all channels (Fz, Cz and Pz). Notably, while our previous study did not examine gamma-band, here we observed significant gamma-band ERD (>30 Hz) across the frontoparietal axis. Gamma-band ERS followed ERD during later post-stimulus periods (600-900 ms).

*NoGo Trial*

The ERSPs observed in the Dutch and Irish cohorts during NoGo trials are presented in Figure 1B. In both centres, ERS was observed in the theta-band and ERD in the alpha- and beta-bands across the frontoparietal axis. Significant beta ERS was also present in the beta-band in both centres. Additionally, gamma-band ERD followed by ERS was observed, most prominently at Cz.

*The difference between NoGo and Go trials*

Figure 2 shows the differences between NoGo and Go trials for both centres. In the Dutch cohort, NoGo trials elicited a more pronounced theta-band ERS at the Fz and Cz electrodes compared with Go trials. Additionally, higher alpha-band ERD and reduced beta- and gamma-band ERS were observed in NoGo trials relative to Go trials. Similar patterns were observed in the Irish cohort.

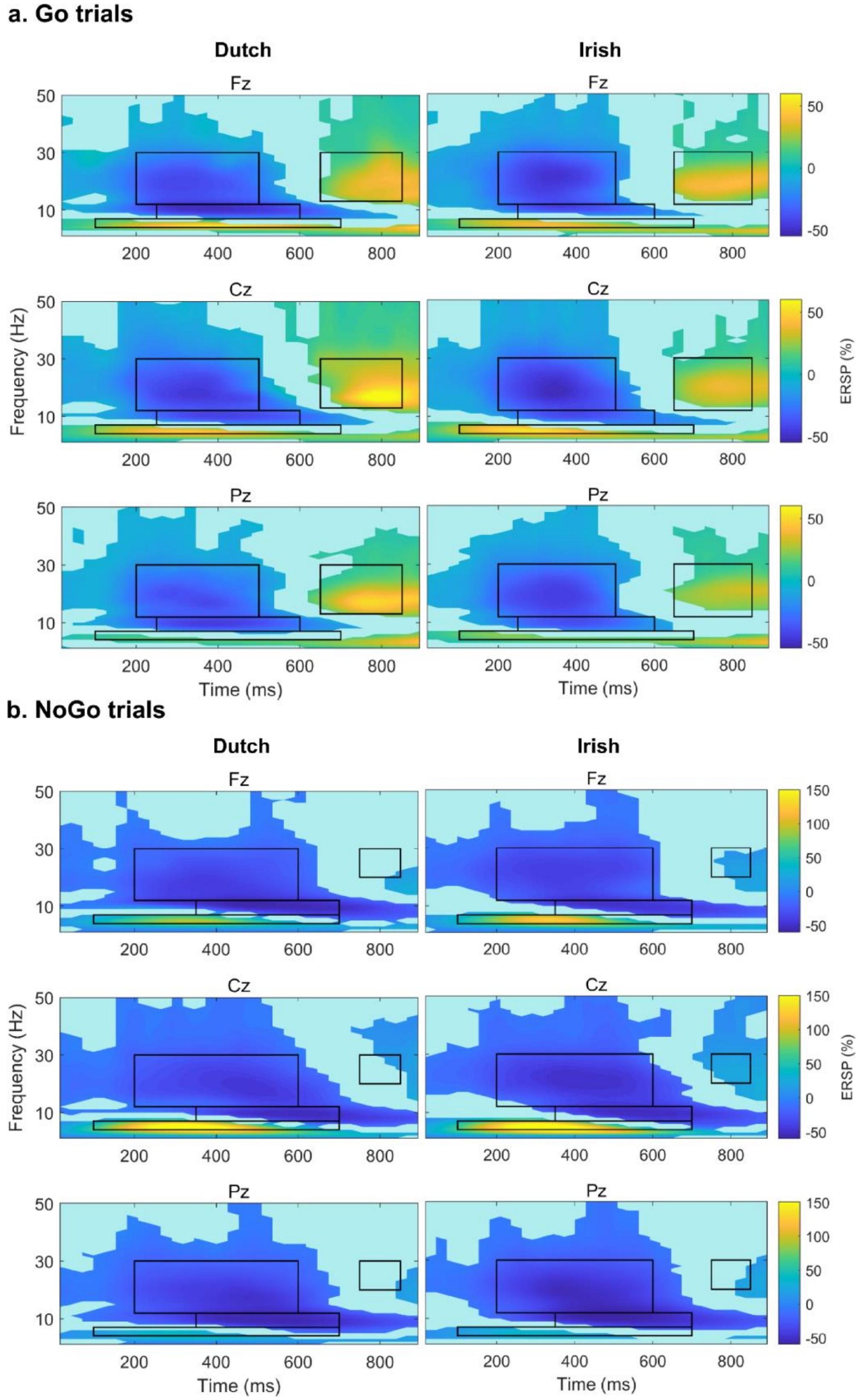

**Figure 1. Significant (a) Go and (b) NoGo trial-related spectral perturbations in controls**. Heatmaps illustrate mean ERSP values for significant effects (sign-rank test, $p_{corrected} < 0.05$). ROIs, defined in our prior study, are outlined with black boxes. Light blue areas indicate no significant spectral perturbation relative to baseline.

Abbreviations: ERSP, event-related spectral perturbation; ROI, region of interest.

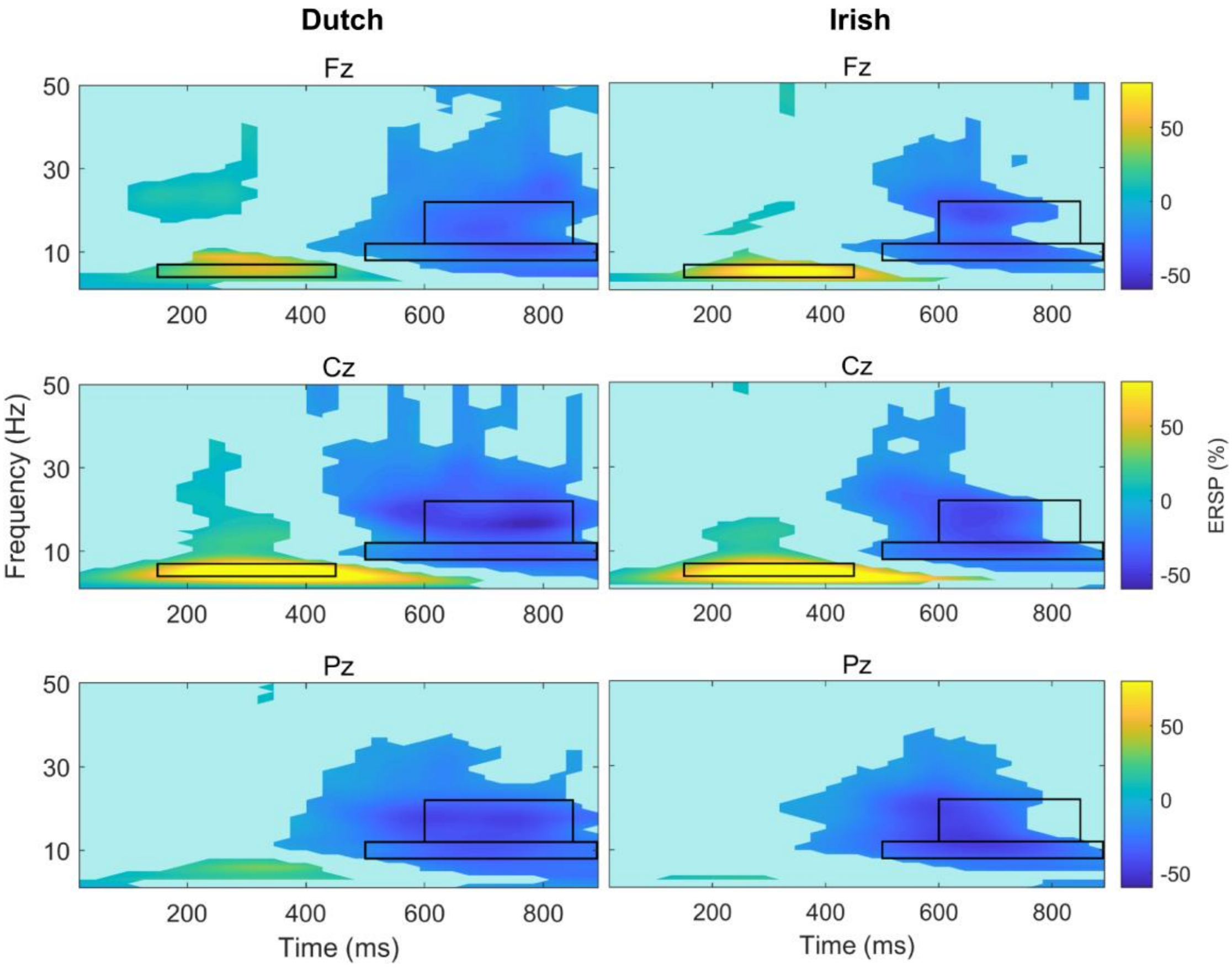

**Figure 2. Significant differences between NoGo and Go trial-related spectral perturbations in controls.** Heatmaps depict mean ERSP differences for significant effects (sign-rank test, $p_{corrected} < 0.05$). ROIs, defined in our prior study, are outlined with black boxes. Light blue areas indicate no significant spectral perturbation relative to baseline.
Abbreviations: ERSP, event-related spectral perturbation; ROI, region of interest.

**3.4 ERSP in ALS compared to controls**

Figure 3 presents the AUROC values for discriminating between people with ALS and controls across the centres, obtained from the Go (Figure 3A) and NoGo trials (Figure 3B).

*Go trial differences in ALS*

For the Dutch cohort during Go trials, no significant difference were observed at any of the analyzed electrodes. In contrast, the Irish cohort revealed several significant regions across electrodes. At Fz, ERS in the theta- and alpha-bands was significantly reduced in ALS (AUROC range = 0.65–0.70). Additionally, beta-band ERD was significantly reduced (i.e. less post-stimulus power decrease) in ALS from 150ms to 400ms, alongside significantly reduced beta-band ERS from 590 ms to 900 ms (AUROC range = 0.65–0.78). We also observed significantly less gamma-band ERD and ERS in those with ALS (AUROC range = 0.65–0.75). At Cz, people with ALS showed significantly reduced theta-band ERS (AUROC range = 0.66–0.71). Additionally, beta-band ERD was significantly reduced from 70 ms to 100 ms and from 230 ms to 370 ms, alongside significantly reduced beta-band ERS from 620 ms to 900 ms (AUROC range = 0.65–0.79). Gamma-band ERD and ERS was also significantly reduced (AUROC range = 0.66–0.73). At Pz, significant effects were confined to the beta-band, with significantly reduced beta-band ERD from 320 ms to 370 ms, and significantly reduced beta-band ERS from 590 ms to 840 ms (AUROC range = 0.71–0.77).

*NoGo trial differences in ALS*

For the Dutch cohort during NoGo trials, no significant regions were observed at any of the analyzed electrodes. In contrast, the Irish cohort showed several significant regions across electrodes. At Fz, people with ALS showed significantly reduced beta-band and gamma-band ERD (AUROC range = 0.69–0.74). At Cz, theta-, alpha-, beta-, and gamma-band ERS were significantly reduced in people with ALS (AUROC range = 0.67–0.74). At Pz, individuals with ALS showed significantly reduced beta- and gamma-band ERD (AUROC range = 0.69–0.77).

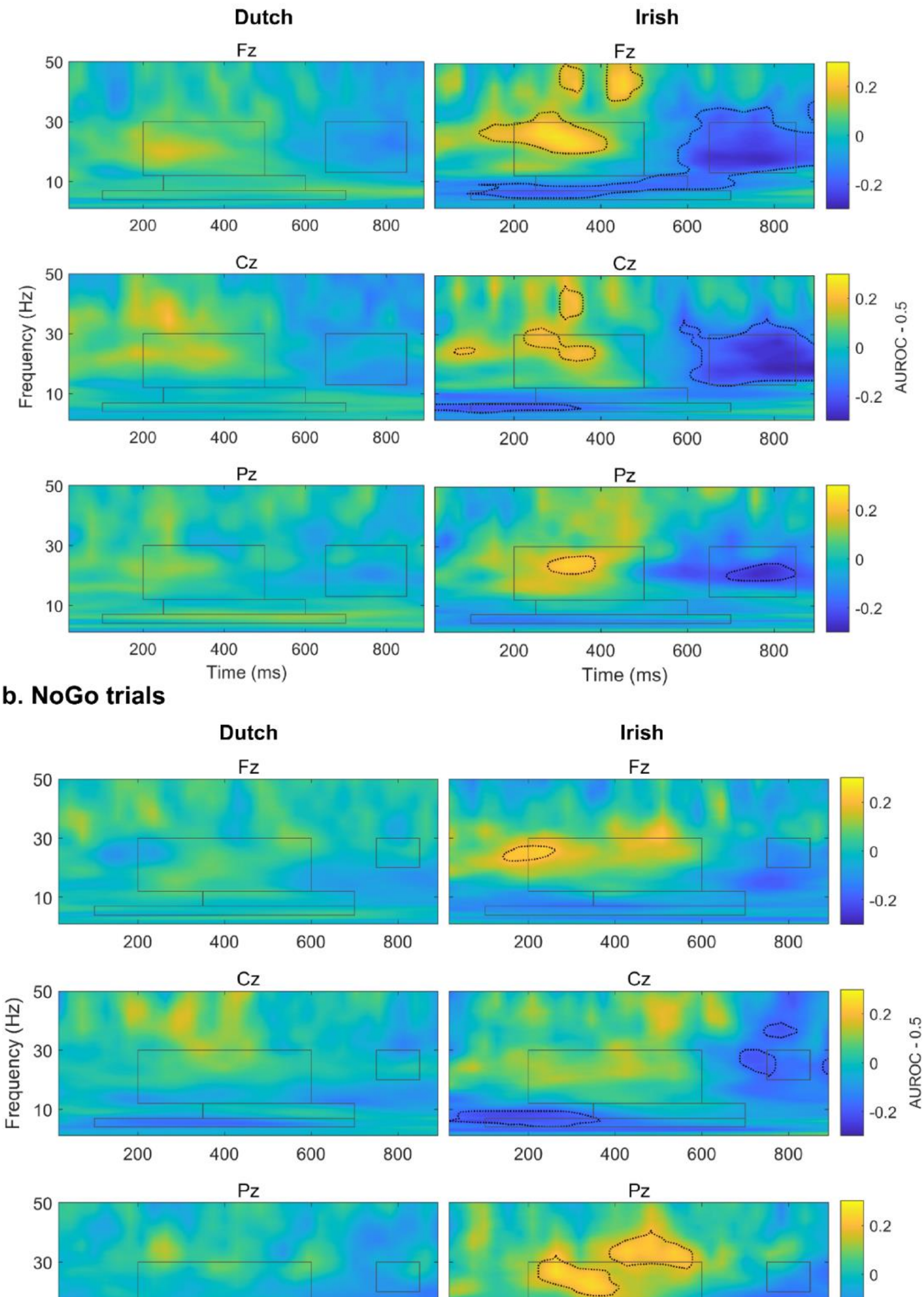
a. Go trials
Dutch
Irish
Fz
Cz
Pz
Frequency (Hz)
Time (ms)
AUROC - 0.5
b. NoGo trials
Dutch
Irish
Fz
Cz
Pz
Frequency (Hz)
Time (ms)
AUROC - 0.5

**Figure 3. Area under the receiver operating characteristic curve minus 0.5 for ERSP changes in ALS compared to controls during (a) Go and (b) NoGo trials, prior to correction for multiple comparisons.** ROIs are outlined with grey boxes. Dotted black lines indicate regions that remain significant after multiple-comparison correction.
Abbreviations: ALS, amyotrophic lateral sclerosis; ERSP, event-related spectral perturbation; ROI, region of interest.

### 3.6 Correlation with task performance

Table 3 presents correlations between beta-band ERS during Go and NoGo trials and task performance measures for both centres, assessing replication of previously reported associations [8]. Table 4 presents correlations between beta-band ERD and task performance measures in the Dutch and Irish ALS cohorts, focusing on task measures that differed between the groups (Go response time, NoGo accuracy, and anticipation error).

*Beta-band ERS*

In the Dutch cohort, negative correlations between Go response time and beta-band ERS during Go trials were observed across all electrodes in both the overall and ALS group, replicating our prior findings. Similarly, in the Irish cohort, such negative correlations were observed across all electrodes in the overall group. In NoGo trials, negative correlations between Go response time and beta-band ERS were observed at Cz in both centres in the overall group, and in the ALS group in the Dutch cohort, again replicating our prior findings. We did not observe the prior reported negative correlation between beta-band ERS during Go trials at Pz and accuracy measures.

*Beta-band ERD*

In the Dutch ALS cohort, negative correlations were observed between beta-band ERD and Go response time during both Go and NoGo trials, as well as a positive correlation with anticipation error at Fz during Go trials. However, none of these correlations remained significant after correction. In the Irish ALS cohort, no significant correlations were observed. This aligns with our prior findings of no significant correlation between beta-band ERD and task performance metrics.

### 3.7 Clinical, demographic, and task-related correlates of ERD and ERS

To further investigate potential cohort differences, we used linear models to determine whether clinical and demographic measures that differed between ALS cohorts were associated with beta-band ERSP measures. Neither King's score, riluzole, nor education reached significance for ERS or ERD at Fz and Pz during Go and NoGo trials. The interaction between King's score and riluzole was not significant for any channel, trial, and measure combination. Finally, a marginal effect of education on ERD at Fz during NoGo trials was identified (F = 3.57, df = 1, p = 0.062), potentially indicating that individuals with lower education display a reduction in ERD by 5.72 units. Detailed results are provided in the Supplementary Material, Section 2.

To further examine potential task-related differences, we investigated whether the number of trials used for time–frequency analysis was associated with beta-band ERSP measures. Significant differences in trial numbers were observed between people with ALS and controls in the Irish cohort (median [interquartile range]: 128 [29] vs. 149 [55], respectively; *p* = 0.01). In addition, the pooled Irish cohort had a higher number of trials than the pooled Dutch cohort (134 [43] vs. 118 [42], respectively; *p* = 0.001).

**Table 3. Correlations between beta-band (13–30 Hz) ERS (%) and SART performance measures**. Negative rho values reflect less ERS with larger behavioral measure value (longer reaction time or greater accuracy). Go trials time window—650–850 ms post stimulus, NoGo trials time window—750–850 ms post stimulus. Uncorrected *p*-values are reported. Bold values indicate results that remain significant after FDR correction (q = 0.05).

| EEG Trial | Channel | Behavior | Group | Dutch | | Irish | |
|---|---|---|---|---|---|---|---|
| | | | | *p* | rho | *p* | rho |
| Go | Fz | Go response time | All | **$1.04*10^{-6}$** | **-0.42** | **$2.27*10^{-5}$** | **-0.48** |
| | | | ALS | **$1.24*10^{-5}$** | **-0.53** | 0.048 | -0.34 |
| | Cz | Go response time | All | **$3.01*10^{-7}$** | **-0.44** | **0.005** | **-0.33** |
| | | | ALS | **$2.73*10^{-4}$** | **-0.45** | 0.640 | -0.08 |
| | Pz | Go response time | All | **$4.57*10^{-4}$** | **-0.31** | **0.008** | **-0.31** |
| | | | ALS | **0.001** | **-0.41** | 0.079 | -0.30 |
| | | Total accuracy | Control | 0.11 | 0.21 | 0.287 | -0.18 |
| | | NoGo accuracy | Control | 0.41 | 0.11 | 0.297 | -0.18 |
| NoGo | Cz | Go response time | All | **$5,75*10^{-05}$** | **-0.35** | **0.011** | **-0.30** |
| | | | ALS | **$6.44*10^{-05}$** | **-0.34** | 0.513 | -0.57 |

Abbreviations: ERS, event-related synchronization; SART, sustained attention to response task; FDR, false discovery rate; ALS, amyotrophic lateral sclerosis; EEG, electroencephalography.

**Table 4. Correlations between beta-band (13–30 Hz) ERD (%) and SART performance measures in people with ALS.** Negative rho values reflect less ERD with larger behavioral measure value (longer reaction time or greater accuracy). Go trials time window—200–500 ms post stimulus, NoGo trials time window—200–600 ms post stimulus. Uncorrected *p*-values are reported. Bold values indicate results that remain significant after FDR correction (q = 0.05).

| EEG Trial | Channel | Behavior | Dutch | | Irish | |
|---|---|---|---|---|---|---|
| | | | ***p*** | **rho** | ***p*** | **rho** |
| Go | Fz | Go response time | 0.12 | -0.20 | 0.65 | 0.08 |
| | | NoGo accuracy | 0.32 | -0.13 | 0.31 | -0.18 |
| | | Anticipation error | 0.04 | 0.26 | 0.32 | 0.17 |
| | Pz | Go response time | 0.02 | -0.29 | 0.76 | 0.05 |
| | | NoGo accuracy | 0.55 | -0.08 | 0.88 | -0.03 |
| | | Anticipation error | 0.19 | 0.17 | 0.23 | 0.21 |
| NoGo | Fz | Go response time | 0.04 | -0.25 | 0.72 | -0.06 |
| | | NoGo accuracy | 0.66 | -0.06 | 0.39 | -0.15 |
| | | Anticipation error | 0.58 | 0.07 | 0.30 | 0.18 |
| | Pz | Go response time | 0.04 | -0.26 | 0.68 | 0.07 |
| | | NoGo accuracy | 0.97 | -0.01 | 0.94 | 0.01 |
| | | Anticipation error | 0.99 | 0.002 | 0.33 | 0.17 |

Abbreviations: ERD, event-related desynchronization; SART, sustained attention to response task; FDR, false discovery rate; ALS, amyotrophic lateral sclerosis; EEG, electroencephalography.

## 4. Discussion

In this replication study, we investigated SART-related cortical oscillatory changes along the frontoparietal axis and examined their relationship with task performance across two independent cohorts from Ireland and the Netherlands. Previously reported SART-related ERSP patterns were replicated in controls across both centres. While prior observed significant differences between people with ALS and controls were replicated in the Irish cohort, this was not replicated in the Dutch cohort.

Similar Go-, NoGo-, and NoGo–Go trial ERSP patterns were observed in controls at both the Irish and Dutch centres, consistent with previous findings [8]. Together, these findings indicate reproducible oscillatory dynamics across independent centres and support the robustness of SART-related oscillatory patterns in controls. Studies using Go/NoGo paradigms have consistently reported ERSP patterns linked to attentional and inhibitory control processes in both healthy and clinical populations. While theta- and alpha-band oscillations are associated with attentional allocation, conflict monitoring, and task engagement [20-22, 26-28], beta-band oscillations are associated with motor processes. Specifically, beta-band ERD and subsequent ERS reflect motor preparation, execution, and response control, with post-movement beta-band ERS often linked to motor cortical inhibition and stabilization of the motor state [23-26].

Alterations in oscillatory activity have been observed across disorders affecting motor and cognitive control. In Parkinson's disease, attenuated post-movement beta-band ERS and altered beta-band ERD within cortico-subcortical motor circuits have been linked to impaired motor and proprioceptive processing [29]. Similarly, altered theta-, alpha-, and beta-band oscillatory responses during Go/NoGo paradigms have been reported in attention deficit hyperactivity disorder and obsessive–compulsive disorder, reflecting disruptions in attentional and inhibitory control [30,31]. Together, these findings highlight the utility of ERSP measures for capturing dysfunction across both motor and cognitive control networks.

Despite the growing body of literature demonstrating altered beta-band ERD and ERS during purely motor paradigms in ALS, the characterization of oscillatory responses in conjunction with cognitive paradigms remains limited. To date, McMackin et al. [8] has characterized SART-related oscillatory activity using defined time-frequency ROIs. The replication of these ERSP patterns in controls across independent centres demonstrates that SART-related oscillatory responses are robust across independent cohorts and recording sites, supporting their use as a neurophysiological reference framework for identifying and interpreting ALS-related alterations.When examining ALS-related alterations, significant regions discriminating between people with ALS and controls were identified in the Irish cohort, replicating the findings reported by McMackin et al. [8]. Consistent with previous work, these regions included reduced beta-band ERD and ERS in people with ALS. In addition, reduced gamma-band activity (not assessed in our prior study) was observed in people with ALS. Gamma-band activity has been associated with attentional engagement and top-down control during sustained task performance [32-34]. The identification of group differences within the gamma band therefore extends the findings of McMackin et al. [8] and may provide additional insight into oscillatory alterations in ALS.

However, these findings were not replicated in the Dutch cohort, where no significant group differences were identified. Given the heterogenous nature of ALS across individuals and disease stages, random population sampling is prone to yielding distinct group pathophysiological profiles. To investigate potential explanations for this discrepancy and examine the utility of these measures as prognostic biomarkers, we examined differences in task performance, as well as demographic and clinical characteristics, between cohorts.

One explanation for the cohort-specific ERS findings relates to differences in response speed between the ALS cohorts. People with ALS in the Irish cohort exhibited slower response times than those in the Dutch cohort. Consistent with our previous work [8], reduced beta-band ERS was associated with slower response times during Go trials across all electrodes and during NoGo trials particularly at Cz. The reduced beta-band ERS observed in the Irish ALS cohort may, therefore, partly reflect the slower response times observed in this group. Alternatively, slower responses may shift the timing of beta-band ERS relative to stimulus onset, thereby influencing stimulus-locked analyses. Together, these findings suggest that beta-band ERS is related to response speed, although the extent to which this reflects underlying motor cortical processing versus temporal shifts in neural activity requires further investigation.

In contrast, associations between beta-band ERD and task performance were inconsistent across cohorts. Although some correlations were observed in the Dutch ALS cohort, these did not survive correction for multiple comparisons, and no associations were observed in the Irish cohort. Consequently, differences in task performance do not readily explain the lack of replication of ERD-related findings.

We also explored whether demographic and clinical differences between centres could account for the discrepant findings. Compared with the Dutch ALS cohort, the Irish cohort showed a higher proportion of individuals in King's stage 1 and stage 4, had higher riluzole use, and a higher educational level. However, exploratory linear models incorporating these variables did not provide evidence that they explained the observed ERSP differences between centres (Supplementary Material, Section 2). Taken together, these findings suggest that the behavioral, demographic, and clinical factors examined in the present study do not fully explain the discrepant findings between centres. Larger, clinically stratified ALS cohorts are needed to replicate these findings and disentangle cohort-specific influences from disease-related effects. Finally, considering the variability of beta-band ERSPs reported in this paper, in future work we will further investigate the stability the diagnostic nature of these metrics in longitudinal studies.

### 4.1 Limitations

Several limitations should be considered when interpreting these findings. We analyzed three key electrodes along the frontoparietal axis so as to replicate our prior examination of ERSP patterns and their relationship with SART performance across Go, NoGo, and Go–NoGo trials. The limited spatial resolution of these findings restricts our ability to attribute ERSP patterns to specific cortical regions, despite capturing activity from key generators involved in motor and cognitive aspects of SART performance. Future studies employing source-localized analyses are needed to identify the cortical sources of the observed oscillatory alterations and potentially enhance the discriminative ability of these measures for detecting ALS. Furthermore, incomplete data collection in the Irish cohort, specifically regarding level of education and cognitive and behavioural profiles, limited our ability to perform robust between-cohort comparisons and may have obscured differences in demographic and clinical characteristics

## 5. Conclusions

Our results in controls are consistent with previously reported SART-related ERSP patterns across independent cohorts, demonstrating reproducible oscillatory dynamics for Go, NoGo, and Go–NoGo contrasts. These findings highlight the value of time-frequency EEG analysis during SART as a complementary approach to event-related potential measures for assessing cognitive and motor network function in ALS. Significant group differences were observed in the Irish cohort but not in the Dutch cohort. Although differences in task performance, demographic characteristics, and clinical factors were explored as potential explanations, these did not fully account for the discrepant findings between centres. The observed association between beta-band ERS and response speed further suggests that these oscillatory measures capture aspects of cortical motor network function during task performance. Together, these findings highlight the potential of SART-related beta-band dynamics as candidate EEG-based markers of motor network dysfunction in ALS, while underscoring the need for further validation in larger, independent, and clinically stratified cohorts.

## Acknowledgements

This study was supported by the Galen and Hilary Weston Foundation (grant number: ...), the Motor Neurone Disease Association UK (grant number: McMackin/Oct20/972-799), and Stichting ALS Nederland, as part of the "GoALS" programme (AV2022-0004, AV2022-0005, AV2023-0006). We would like to thank all the people with ALS, participants and their families who volunteered to take part in this study. We thank the Wellcome-HRB Clinical Research Facility at St. James's Hospital for providing a dedicated environment for the conduct of high-quality clinical research.

# Supplementary Materials - Beta-band oscillation changes in ALS: A Dual-Site International Replication Study

Section 1 introduces the receiver operating characteristic (ROC) curve analysis, presenting the sensitivity and specificity values obtained for both cohorts. Section 2 further examines the relationship between event-related desynchronization (ERD)/ synchronization (ERS) and ALS-related features, including King's staging, country of cohort, education level, and riluzole use. Finally, in Section 3, we include additional details about the preprocessing validation.

## 1. ROC Curve Analysis

Five regions of interest (ROIs) were selected based on consistent findings across previous and current work, which showed significant AUC deviations in the Irish cohort during Go and NoGo trials in the beta-band band (13-30 Hz). The selected ROIs were NoGo beta-band-band ERD (200-600 ms), Go beta-band ERD (200-500 ms) at both Fz and Pz, and Go beta-band ERS (650-850 ms) at Fz only.

Mean ITV (without downsampling) was calculated within the time–frequency ranges of the selected ROIs. Feature values were inverted when necessary to ensure that higher values consistently corresponded to the patient class. Discriminative performance between patients and controls was assessed separately for the Irish and Dutch cohorts using ROC analysis. The area under the ROC curve (AUC) was computed using the Mann–Whitney U statistic. ROC curves were constructed by evaluating all unique feature values as decision thresholds, classifying observations as positive when the feature value was greater than or equal to the threshold, and computing the corresponding sensitivity and specificity at each threshold. The optimal operating point was defined using Youden's index (J = sensitivity + specificity − 1), and the corresponding optimal threshold, sensitivity, and specificity were reported. Confidence intervals for the AUC were estimated using DeLong's method [1].

Supplementary Figure 1 shows the ROC curves of beta-band ERD and ERS at Fz during Go and NoGo trials for both centres. In the Irish cohort, moderate discriminative performance was observed for NoGo beta-band ERD (AUC = 0.620, 95% CI: 0.486–0.753) and Go beta-band ERD (AUC = 0.644, 95% CI: 0.513–0.774). The highest discriminative performance was observed for Go beta-band ERS (AUC = 0.780, 95% CI: 0.672–0.889), indicating good separation between patients and controls. At the optimal operating point, Go beta-band ERS achieved a sensitivity of 0.81 and specificity of 0.69, whereas NoGo beta-band ERD and Go beta-band ERD showed sensitivities/specificities of 0.67/0.64 and 0.64/0.67, respectively. In the Dutch cohort, NoGo beta-band ERD demonstrated low discriminative ability (AUC = 0.541, 95% CI: 0.439–0.643), with high sensitivity (0.95) but poor specificity (0.19) at the optimal threshold. Go beta-band ERD showed moderate discriminative performance (AUC = 0.632, 95% CI: 0.533–0.730), with sensitivity and specificity values of 0.52 and 0.73, respectively. Go beta-band ERS also demonstrated moderate discriminative performance (AUC = 0.620, 95% CI: 0.521–0.720), yielding a sensitivity of 0.51 and specificity of 0.83 at the optimal operating point.

Supplementary Figure 2 shows the ROC curves of beta-band ERD at Pz during Go and NoGo trials for both centres. In the Irish cohort, moderate discriminative performance was observed for NoGo beta-band ERD (AUC = 0.686, 95% CI: 0.560–0.811) and Go beta-band ERD (AUC = 0.662, 95% CI: 0.536–0.788). At the optimal operating point, NoGo beta-band ERD achieved a sensitivity of 0.58 and specificity of 0.78, whereas Go beta-band ERD showed a sensitivity of 0.72 and specificity of 0.56. In the Dutch cohort, both NoGo and Go beta-band ERD demonstrated poor discriminative ability (AUC = 0.531, 95% CI: 0.429–0.634 and AUC = 0.553, 95% CI: 0.452–0.654, respectively). At the optimal operating point, NoGo beta-band ERD showed high sensitivity (0.81) but low specificity (0.33), while Go beta-band ERD demonstrated a sensitivity of 0.73 and specificity of 0.40.

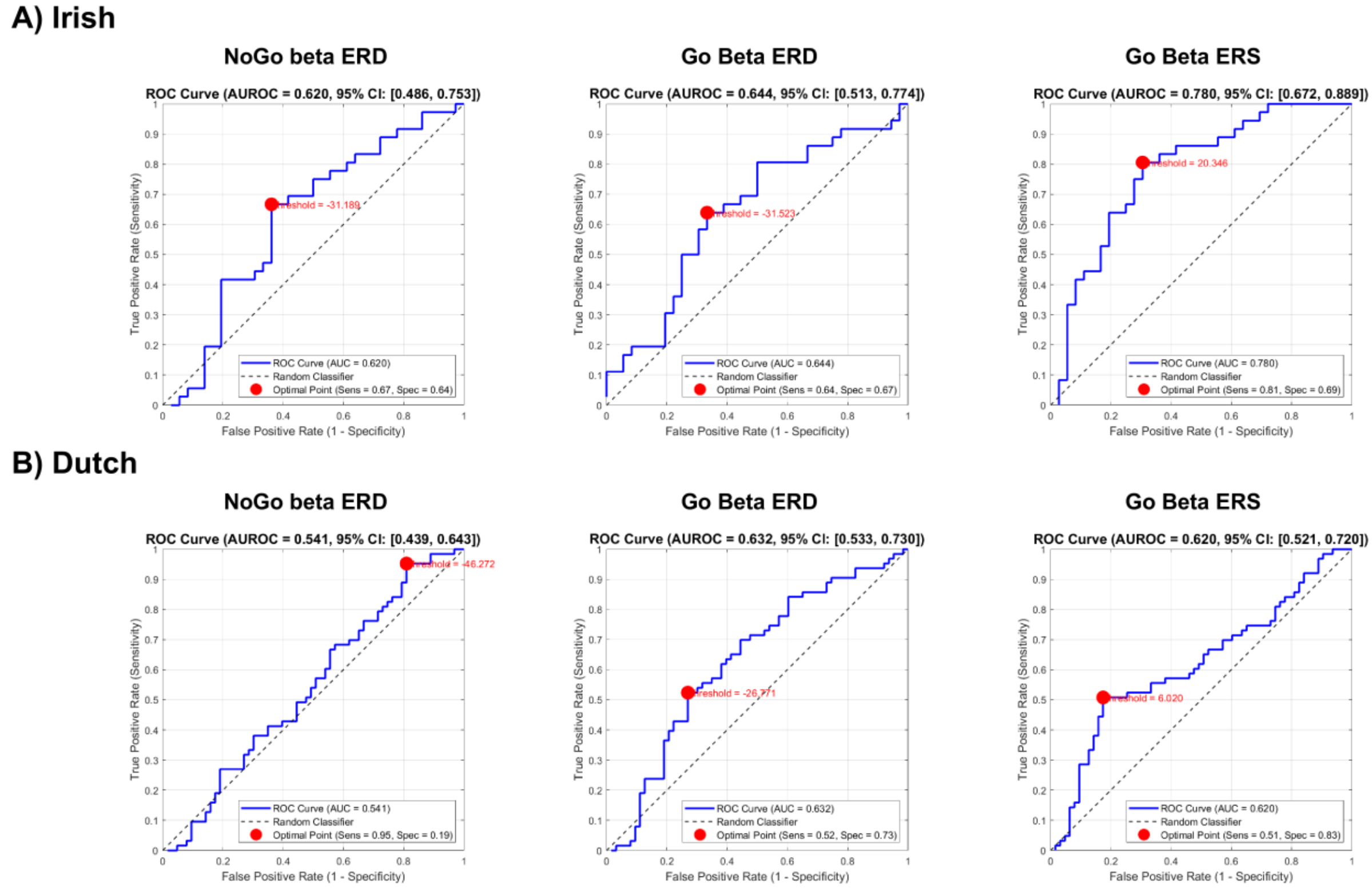

**Supplementary Figure 1: ROC curves of beta-band ERD and ERS (13-30 Hz) at Fz during Go and NoGo trials.** Time windows were defined as 200-600 ms (NoGo ERD), 200–500 ms (Go ERD) and 650–850 ms (Go ERS) post-stimulus.

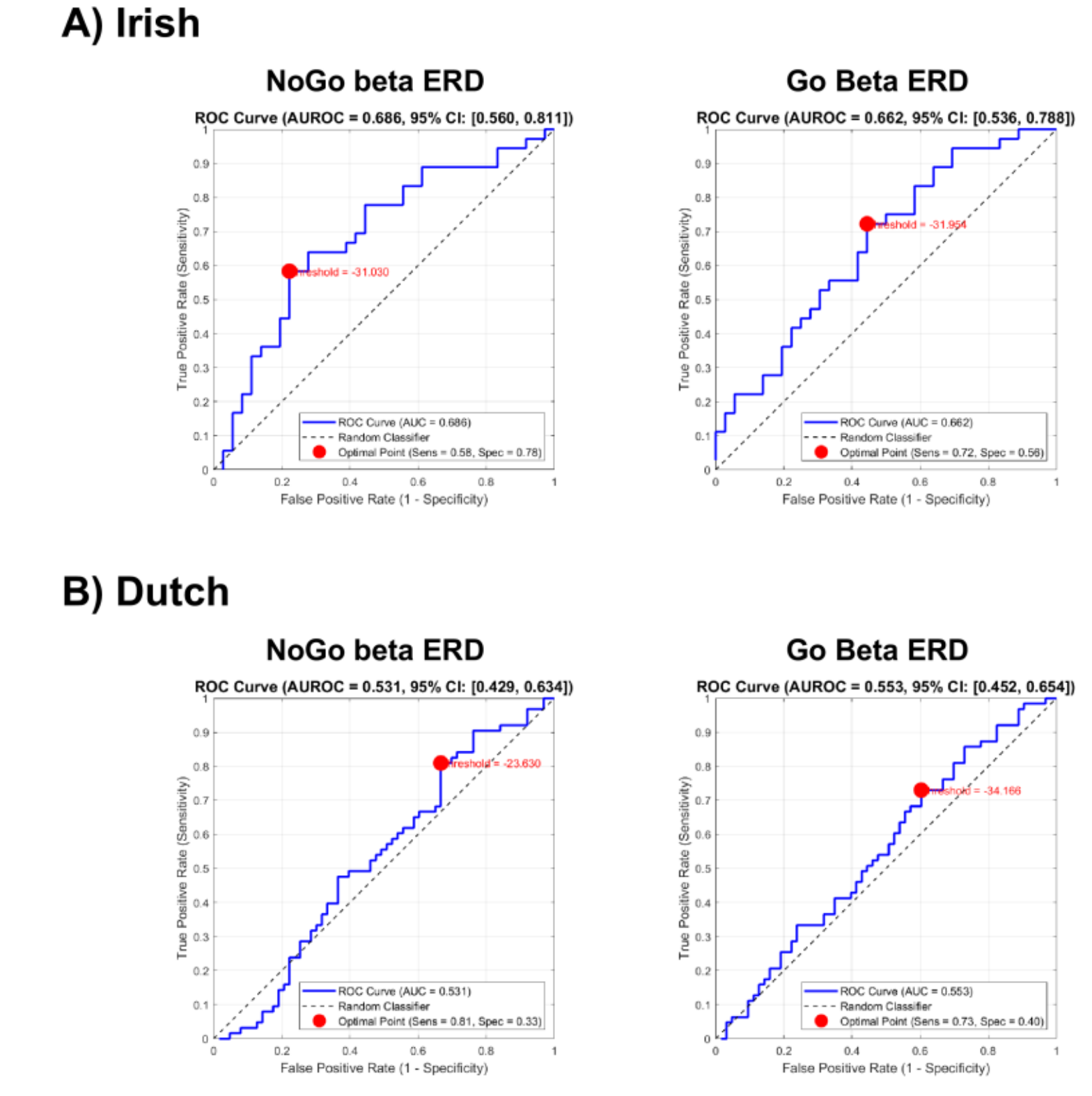

**Supplementary Figure 2: ROC curves of beta-band ERD at Pz during Go and NoGo trials.** Time windows were defined as 200–500 ms (Go ERD), and 200–600 ms (NoGo ERD) post-stimulus.

## 2. Clinical and demographic correlates of ERD and ERS

For ERS at Fz during Go trials, the interaction between King's score and riluzole was not significant (F = 1.09, df = 3, p = 0.36). Also, country was significant (F = 7.40, df = 1, p = 0.008), while King's score (F = 0.15, df = 3, p = 0.93), riluzole (F = 0.01, df = 1, p = 0.91) and education (F = 0.64, df = 1, p = 0.43) were not significant. For ERD at Fz during Go trials, the King's score × riluzole interaction was not significant (F = 0.87, df = 3, p = 0.46), and no predictor reached significance: King's score (F = 0.80, df = 3, p = 0.50), country (F = 0.03, df = 1, p = 0.87), riluzole (F = 0.23, df = 1, p = 0.64) and education (F = 1.09, df = 1, p = 0.30).

For ERS at Fz during NoGo trials, the King's score × riluzole interaction was not significant (F = 0.58, df = 3, p = 0.63). Also, no predictor reached significance: King's score (F = 0.56, df = 3, p = 0.64), country (F = 0.34, df = 1, p = 0.56), riluzole (F < 0.01, df = 1, p = 0.98) and education (F = 0.13, df = 1, p = 0.72). For ERD at Fz during NoGo trials, the King's score × riluzole interaction was not significant (F = 1.46, df = 3, p = 0.23). Finally, education showed a marginal effect (F = 3.57, df = 1, p = 0.062), while King's score (F = 0.09, df = 3, p = 0.96), country (F = 1.96, df = 1, p = 0.17) and riluzole (F = 0.18, df = 1, p = 0.67) were not significant.

For ERS at Pz during Go trials, the King's score × riluzole interaction was not significant (F = 0.88, df = 3, p = 0.46). Country was significant (F = 7.94, df = 1, p = 0.006), whereas King's score (F = 0.27, df = 3, p = 0.85), riluzole (F = 0.70, df = 1, p = 0.41) and education (F = 0.002, df = 1, p = 0.96) were not significant. For ERD at Pz during Go trials, the King's score × riluzole interaction was not significant (F = 0.37, df = 3, p = 0.78), and no predictor reached significance: King's score (F = 0.56, df = 3, p = 0.64), country (F = 1.03, df = 1, p = 0.31), riluzole (F = 0.51, df = 1, p = 0.48) and education (F = 0.05, df = 1, p = 0.82).

For ERS at Pz during NoGo trials, the King's score × riluzole interaction was not significant (F = 1.55, df = 3, p = 0.21). Also, no predictor reached significance: King's score (F = 0.64, df = 3, p = 0.59), country (F = 1.31, df = 1, p = 0.26), riluzole (F = 0.06, df = 1, p = 0.81) and education (F = 0.10, df = 1, p = 0.76). For ERD at Pz during NoGo trials, the King's score × riluzole interaction was not significant (F = 1.50, df = 3, p = 0.22). Country was significant (F = 4.10, df = 1, p = 0.046), while King's score (F = 2.08, df = 3, p = 0.11), riluzole (F = 0.07, df = 1, p = 0.79) and education (F = 0.61, df = 1, p = 0.44) were not significant.

## 3. Preprocessing Validation

To compare the proposed automated preprocessing pipeline with the pipeline used by McMackin et al. [1], ERP signals obtained during Go and NoGo trials were visually inspected for all participants. Supplementary Figure 3 presents the overall ERP waveforms generated by the automated pipeline alongside those obtained using the pipeline described by McMackin et al. [2].

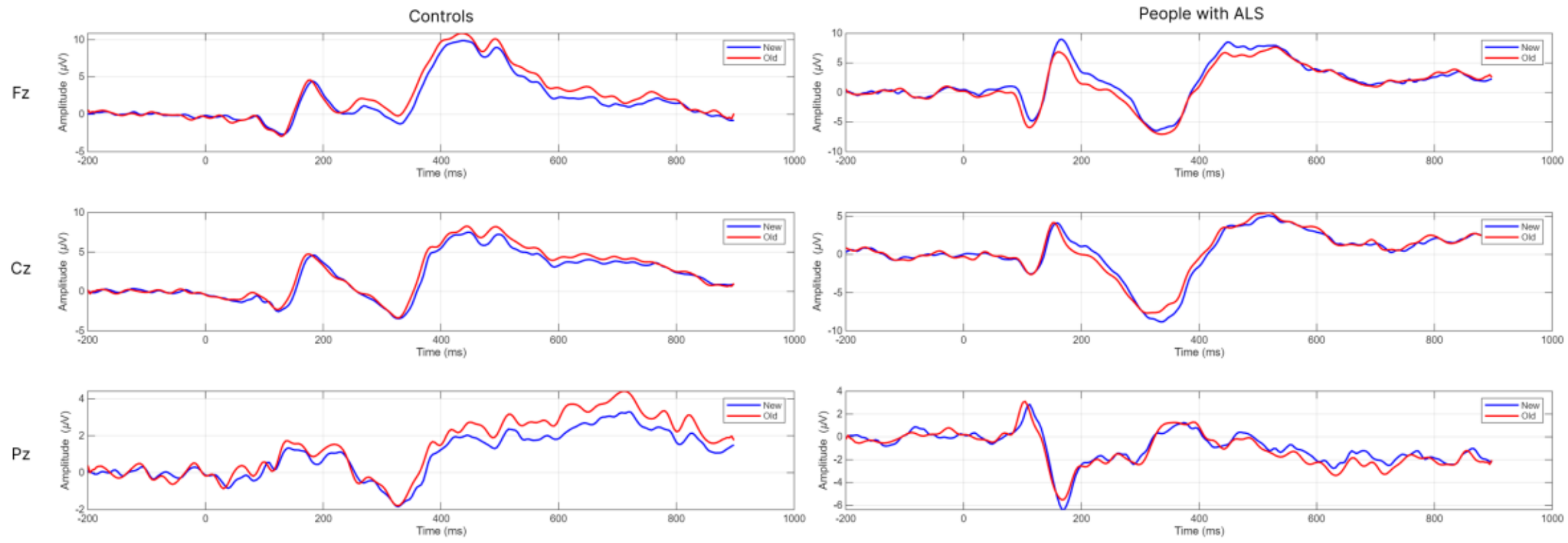

**Supplementary Figure 3: ERP signals for the three channels during NoGo trials for one control and one individual with ALS.**

Finally, N2 and P3 peak amplitudes and latencies were compared between pipelines. For the N2 component, peak amplitudes were strongly positive correlated across pipelines for both No-Go trials (Fz: ρ = 0.95, Cz: ρ = 0.95, Pz: ρ = 0.88; all p < 0.01) and Go trials (Fz: ρ = 0.97, Cz: ρ = 0.98, Pz: ρ = 0.89; all p < 0.01). N2 latency correlations were slightly lower, ranging from ρ = 0.81 to 0.87 for No-Go trials and from ρ = 0.66 to 0.78 for Go trials (all p < 0.01).

For the P3 component, peak amplitudes were also strongly positive correlated across pipelines for both No-Go trials (Fz: $\rho = 0.89$, Cz: $\rho = 0.97$, Pz: $\rho = 0.90$; all $p < 0.01$) and Go trials (Fz: $\rho = 0.92$, Cz: $\rho = 0.96$, Pz: $\rho = 0.96$; all $p < 0.01$). P3 latency correlations ranged from $\rho = 0.85$ to 0.94 for No-Go trials and from $\rho = 0.68$ to 0.93 for Go trials (all $p < 0.01$). Overall, the two pipelines produced similar ERP signals. Given the advantages of the automated pipeline for ensuring reproducibility across both the Dutch and Irish centres, the automated pipeline was used for all analyses.